\documentclass[a4paper]{article}
\usepackage[margin=1in]{geometry} 

\usepackage{amsmath}
\usepackage{amsthm}
\usepackage{amssymb}

\usepackage{graphicx}
\usepackage{subfig}
\usepackage{upgreek}
\usepackage{color,soul}

\usepackage[T2A]{fontenc}
\usepackage[utf8]{inputenc}
\usepackage[english]{babel}
\usepackage{hyperref}

\usepackage[sort&compress,numbers,square]{natbib}
\usepackage{verbatim}

\begin{document}
\noindent\textbf{\Large{Multiple spiking functionalities in annealing-optimized\\ Ag/Hf$_{0.5}$Zr$_{0.5}$O$_{2}$-based memristive neurons}}

\vspace{5mm}\noindent\large{Nikita Zhidkov, Andrei Zenkevich, Anton Khanas$^*$}

\vspace{3mm}\noindent\normalsize{\textit{Lab of Functional Materials and Devices for Nanoelectronics, Moscow Institute of Physics and Techno-\\logy (National Research University), Institutsky lane 9, Dolgoprudny, Moscow region, 141700, Russian Federation}}

\vspace{3mm}\noindent$^*$Corresponding author: \texttt{khanas@phystech.edu}

\section*{Abstract}

Rapid progress of artificial neural network applications in recent years has led to the issue of an unprecedented energy consumption. It can be solved by the implementation of energy efficient hardware based on non-von-Neumann architectures, which requires the development of electronic components emulating the behavior of synapses and neurons. While research of synaptic elements is vast, the technology for fabrication of scalable and highly reproducible neuronal elements is far less developed. In this paper, we demonstrate an artificial neuron with multiple functionalities based on filamentary switching Ag/Hf$_{0.5}$Zr$_{0.5}$O$_{2}$ (HZO) memristors. To improve the parameters of memristors, we propose a two-step annealing method, which allows for better control of the crystallization of the functional dielectric layer (HZO) as well as of the diffusion of active electrode (Ag) atoms. Furthermore, we demonstrate the leaky integrate-and-fire (LIF) neuronal behavior in multiple spiking modes: time-to-first-spike (TTFS), number of spikes and firing rate coding. Moreover, the neuron operation does not require the additional electronic overhead and is supported solely by a Ag/HZO memristor with a current limiting resistor connected in series. The presented results pave the way for the creation of next generation energy efficient neuromorphic hardware operating on the principles of spiking neural networks.

\vspace{12pt}{\noindent}\textbf{Keywords:} memristor, hafnium-zirconium oxide, Ag ion filaments, leaky integrate-and-fire, spiking neuron, time-to-first-spike coding

\newpage
Massive parallelism, low energy consumption and high scalability make neuromorphic computing architecture a promising candidate for replacing von Neumann architecture in the hardware for artificial intelligence \cite{schuman_opportunities_2022}. In order to enable the full capabilities of neuromorphic computing, the dedicated hardware has to be built from the elements emulating behavior of brain synapses and neurons on the physical level \cite{zhang_neuro-inspired_2020}. Memristors are among the most promising microelectronic circuit elements for the implementation of both electronic synapses \cite{aguirre_hardware_2024} and neurons \cite{liu_artificial_2023}.

{\noindent}Memristors are two-terminal devices that can change their resistance under electrical stimuli and retain its changed state for some time \cite{strukov_missing_2008, kumar_dynamical_2022}. Non-volatile memristors can be used as artificial synapses when there is an option of continuous resistance state tuning, leading to various synaptic plasticity effects \cite{zhu_comprehensive_2020}. Due to high promise for physical realization of vector-matrix multiplication, memristive synapse research has been in the spotlight of the community over the last decade \cite{aguirre_hardware_2024}. However, the implementations of memristive artificial neurons has been significantly less explored, despite being equally important in the context of realization of spiking neural networks \cite{zuo_volatile_2023}. In order to create an electronic artificial neuron, one has to implement the physical elements and to connect them according to a chosen neuron model. The models differ by the number and sophistication of their elements, which correlates with the degree of biological complexity that can be emulated \cite{izhikevich_which_2004}. The necessary constituents of an electronic artificial neuron can be described on the example of the simplest spiking neuron model - leaky integrate-and-fire (LIF) \cite{gerstner_neuronal_2014}. An equivalent circuit of a LIF neuron comprises three elements: a membrane capacitance, a leakage resistance and a threshold switch (TS). Their biomimetic roles are as follows: a membrane capacitance accumulates ("integrate") electric charge from the neuron inputs; a leakage resistance ensures the decay of electric charge when the neuron is not activated ("leaky"); and a TS realizes the mechanism spike generation at the output terminal (axon) in case the accumulated electric charge exceeds a certain threshold value ("fire"). Since in terms of dynamical systems this also describes a self-oscillator, a necessary condition for neuron functionality is the presence of negative differential resistance (NDR) region in the $I(V)$ curve of a TS \cite{kumar_dynamical_2022}. Development of appropriate TS is at the forefront of this field, since other two elements can be easily implemented by capacitors and resistors.

{\noindent}There is a large variety of physical mechanisms and materials that can be used for TS creation \cite{han2022}: metal-insulator transition \cite{pickett_scalable_2013, yi2018biological}, ovonic switching \cite{lee_various_2019}, single-transistor latching \cite{han_leaky_2018}, gas discharge ignition \cite{trunov_2025}. One of the most widespread implementations is based on volatile filamentary switching memristors \cite{han2022}. Filament is a localized conducting path inside the functional dielectric layer connecting the electrodes of a memristor \cite{woo_localized_2025}. Due to the small size of the filament (of the order of $1\div10~$nm), the filamentary memristors have great potential for creation of ultra-scaled elements for neuromorphic chips with high element density. In order to ensure the low power consumption and high signal-to-noise ratio during spike generation, it is required to create TS with low threshold voltage ($V_\text{th}$) and high $I_\text{on}/I_\text{off}$ ratio, while preserving the resistance state bistability property -- which in practice is equivalent to NDR \cite{lee_various_2019}. These demands justify the utilization of "active" metal electrodes as a source of material for volatile filament formation (electrochemical metallization, ECM) -- such as Ag or Cu protruding a dielectric layer under the voltage application (\cite{wang_fully_2018, zhou_thermally_2023}). Additional functionality or reconfigurability of memristive neurons is also desirable, since it would allow to create flexible and multi-purpose neuromorphic devices. In particular, multiple neuronal functionalities in one device \cite{yi2018biological, zhao_spiking_2025}, electrical \cite{lewerenz_threeterminal_2024} or light \cite{li_crossmodal_2024} tunability of membrane capacitance or leakage resistance by additional memcapacitors \cite{ignatov_memristive_2015, feali_implementation_2018} or memristors \cite{xiao_bio-plausible_2025} are among such supplementary features.
    
{\noindent}One of the main issues in filamentary TS operation is the cycle-to-cycle variability of switching parameters, such as $V_{th}$, $I_\text{on}$ and $I_\text{off}$. Many solutions to this problem have been proposed: fabricating Ag-based nano-objects (via annealing, lithography or template deposition) to regulate the morphology and limit the amount of Ag for filament formation \cite{shukla_ag_hfO2_based_2016, li_highuniformity_2020, hua_ag/hfo2-based_2021, long_effects_2023}; inducing a diffusion barrier to control injection of Ag atoms into a dielectric during electroforming \cite{grisafe_performance_2019, ke_highly_2024}; using carbon nanotube and Ag nanowire as electrodes to spatially confine the switching region and induce self-compliant behavior \cite{hu_selfcompliant_2025}. However, the proposed methods, despite solving some of the challenges, involve processes that are either difficult to scale or that increase the complexity of the fabrication route. Thermal annealing is a method that is free from these drawbacks. Two main phenomena that occur in the active metal (Ag, Cu)/functional dielectric heterostructures during thermal annealing are crystallization of a dielectric layer and diffusion of active metal atoms into the dielectric. The latter occurs mainly through the grain boundaries (GB) \cite{mishin_grain_1997} and triple junctions (TJ) \cite{petelin_triple_2001}, formed during crystallization, and strongly influences the resistance switching properties of the device \cite{hua_lowvoltage_2019}. The crystallization of a dielectric layer has also been shown to have a drastic effect \cite{lee_crystallinitycontrolled_2025, guo_effect_2019}. Due to the simultaneous flow of these processes, it is difficult to control the result, nevertheless, careful optimization of annealing steps is a promising route in improvement of TS properties.

{\noindent}In this work, we demonstrate a Ag/Hf$_{0.5}$Zr$_{0.5}$O$_{2}$ (HZO) memristive neuron with multiple spiking functionalities. First, we propose an optimized procedure for TS fabrication based on the introduction of two thermal annealing steps, which allows to obtain memristors with $V_\text{th}$ as low as 0.5$~$V, $I_\text{on}/I_\text{off}$ ratio as high as $10^6$ and steep single-step switching. Furthermore, we demonstrate a LIF neuron with the possibility of operation in three different modes of information coding, controlled by the value of input voltage: time-to-first-spike, number of spikes and firing rate coding. We believe that the demonstrated effects, both the optimization of TS fabrication flow and creation of a spiking neuron with multiple functionalities, constitute an important step towards development of energy efficient neuromorphic chips.


\begin{figure}[t!]
    \centering
    \includegraphics[width=\linewidth]{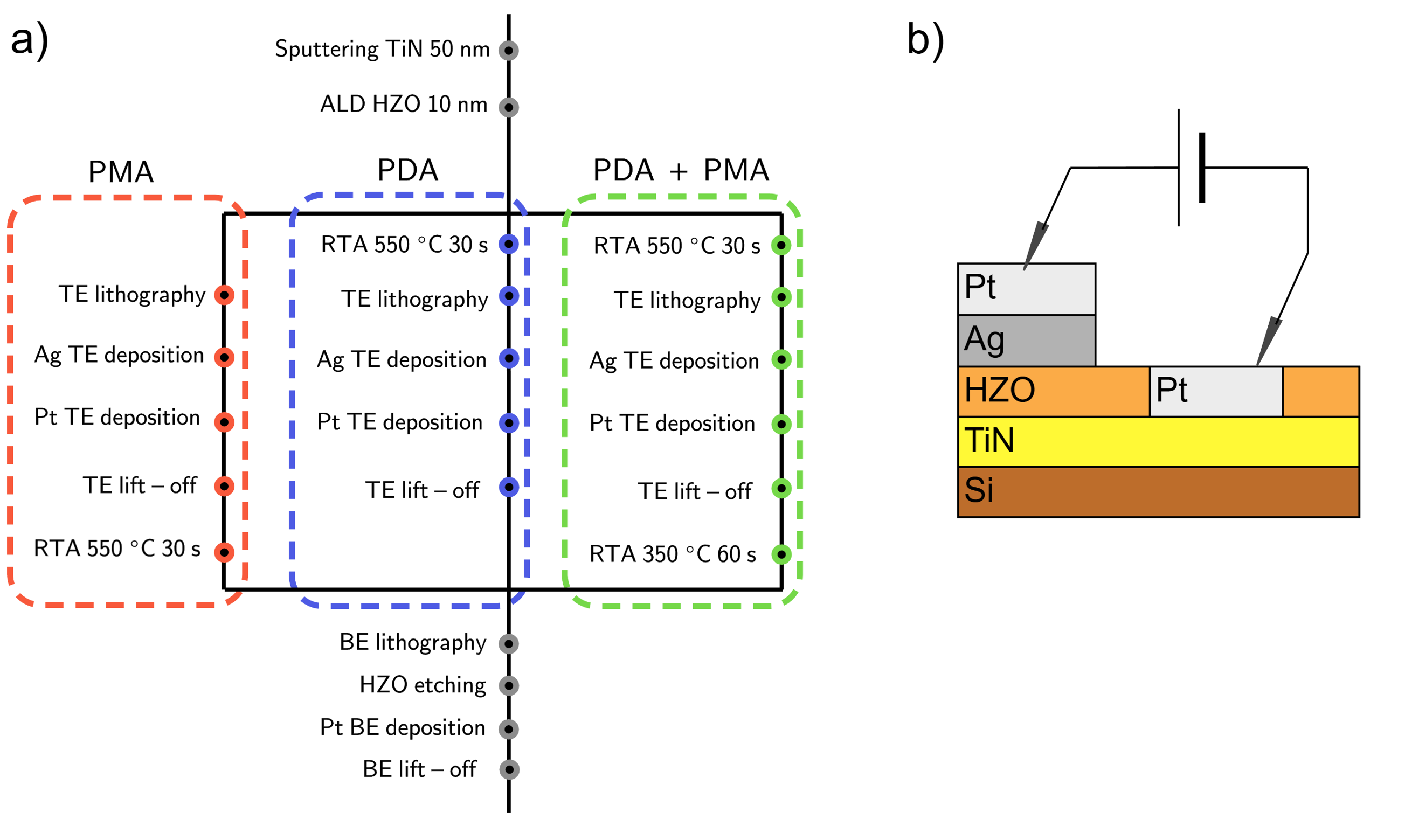}
    \caption{a) The fabrication process flow and b) final structure sketch with connections for electrical measurements.}
    \label{fig:samples}
\end{figure}
    
\textit{Device fabrication:} The fabrication process flow of Ag/HZO-based memristors is shown in \textbf{fig.\ref{fig:samples}a}. First, a 50 nm TiN layer, acting as a bottom electrode (BE), was deposited on a Si(100) substrate via DC magnetron sputtering. Then, a 10-nm-thick HZO layer was grown by atomic layer deposition (ALD), following a standard procedure described elsewhere \cite{zarubin_fully_2016, chouprik_effect_2023}. Next, the shape of top electrode (TE) was defined via photolithography, after which a Ag TE layer with a Pt protecting overlayer were deposited and lifted-off. Ag layer was fabricated using two techniques: pulsed laser deposition (PLD) and electron beam (e-beam) evaporation, which yielded similar results. PLD was carried out using a Q-switched Nd:YAG laser (1064 nm wavelength) with $\sim$10$^8~$W$~$cm$^{-2}$ power density and 10 Hz pulse repetition rate (energy per pulse $\sim45~$mJ) in vacuum $\le 5\cdot10^{-7}~$mbar. During e-beam evaporation the pressure in the vacuum chamber was $\le10^{-6}~$mbar, the cathode current was set to 140$~$mA and the deposition rate equaled 0.5 \AA/s. To establish reliable contact to the BE, another lithography step was conducted, followed by HZO plasma etching, Pt deposition atop of TiN and lift-off. Heat treatment was performed by rapid thermal annealing (RTA) in two options: at $550~^{\circ}$C during 30 seconds and at $350~^{\circ}$C during 60 seconds -- at various steps of the fabrication process (described in more detail below). 

{\noindent}\textit{Electrical measurements:} Keysight B1500A semiconductor device analyzer was used for all electrical measurements. During all measurements, the BE was grounded, while potential was applied to the TE.
The connection scheme is shown in \textbf{fig\ref{fig:samples}b}. In the neuronal spiking experiments, the series resistance $R_{\text{s}}=2.7~$M$\Omega$ was connected to the device externally to limit the maximum current.



In order to obtain the Ag/HZO/TiN memristors with the properties optimized for the spiking neuron implementation, steps of heat treatment were introduced at different stages of the fabrication process. We note that in all our experiments the fabrication route without RTA steps yielded memristors with non-volatile switching (not shown) in agreement with other reports \cite{lee_crystallinitycontrolled_2025}, which was not suitable for our goal. The annealing of our samples manifests in two main phenomena: crystallization of HZO and diffusion of Ag inside HZO, both of which can have drastic effect on the electrical properties of the memristors.

\begin{figure}[t!]
    \centering
    \includegraphics[width=\linewidth]{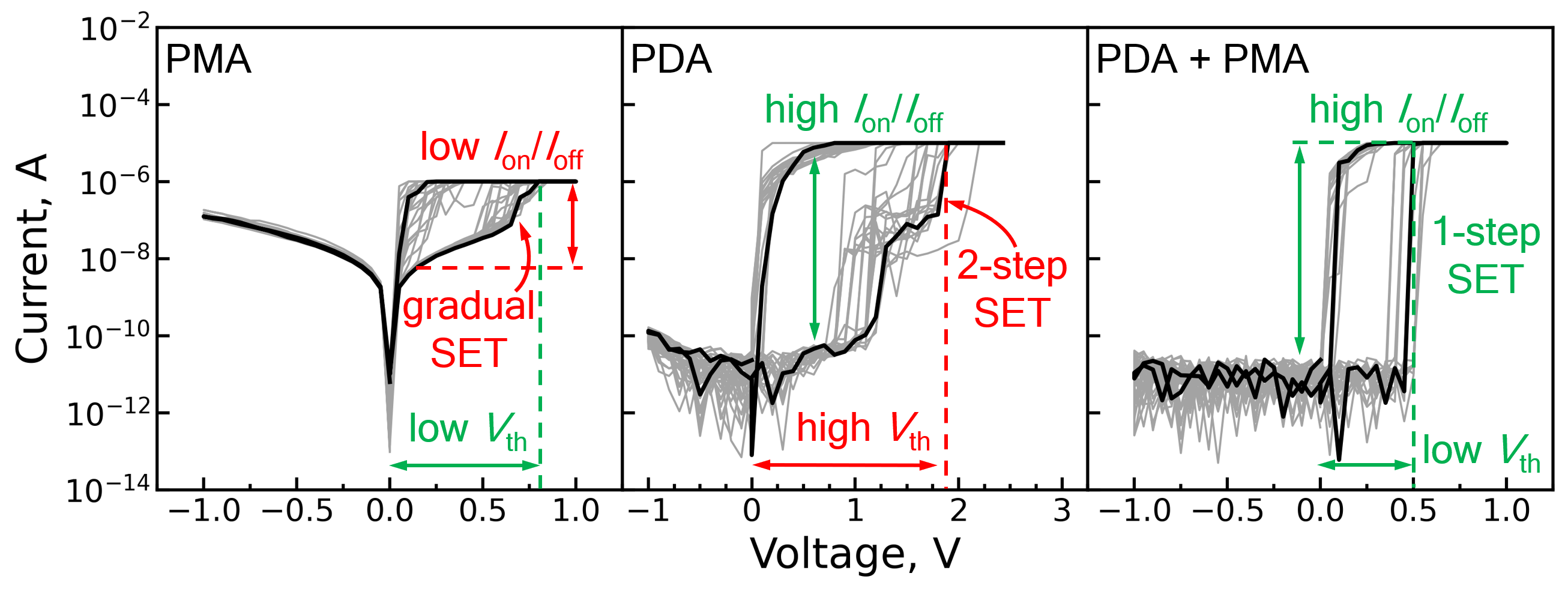}
    \caption{$I(V)$ curves over 20 cycles each (grey lines; a representative cycle - black line), demonstrating threshold switching in memristors after different production flows: only post-metallization annealing (PMA), only post-deposition annealing (PDA) and both PDA and PMA.}
    \label{fig:IV}
\end{figure}

{\noindent}\textbf{Fig.\ref{fig:IV}} illustrates the $I(V)$ curves measured on the devices obtained after three heat treatment configurations. In the first option, RTA was performed after Ag deposition (post-metallization annealing, PMA). After this process amendment we already obtain volatile threshold switching instead of non-volatile switching. The $I(V)$ curves are characterized by low threshold voltages ($V_\text{th} \approx 0.7~$V), which is an advantage in terms of potential low power consumption applications. On the other hand, we observe low $I_\text{on}/I_\text{off} \approx 17$ ratio due to the low resistance in the OFF state ($R_\text{off}$) and the necessity for low current compliance ($I_\text{c.c.}$) level. We believe that in the case of PMA at high temperature, crystallization of HZO and diffusion of Ag inside HZO occur simultaneously. Consequently, Ag atoms diffuse deep inside the HZO layer, reducing the value of $R_\text{off}$ down to $ \approx 8 \cdot 10^6~\Omega$. Additionally, to stabilize the switching of the devices and prevent them from early breakdown, the $I_\text{c.c.}$ had to be set at $10^{-6}~$A, which further reduces the value of $I_\text{on}/I_\text{off}$, critical for the efficient neuromorphic operations. Moreover, it can be seen that the PMA devices demonstrate mostly gradual threshold switching, which is also unfavorable for artificial neuron implementation, since it does not allow to realize resistance state bistability. Thus, it was concluded that while PMA treatment yields some improvement in the device properties, it is still not optimal for practical implementations.

{\noindent}In the following series of devices, RTA was performed only after the growth of HZO (post-deposition annealing, PDA). In this case, the $I(V)$ curves show average the $I_\text{on}/I_\text{off} \approx 2 \cdot 10^5$, which is a substantial improvement over the PMA case. This is due to the higher $R_\text{off} \approx 5 \cdot 10^{10}~\Omega$ and more stable switching character, which allowed to set current compliance up to $I_\text{c.c.} = 10^{-5}~$A. However, we observe significantly higher threshold voltages $V_\text{th} \approx 1.1~$V, which is a disadvantage of this method. Moreover, SET (switching from OFF to ON state) occurs in two steps, sometimes, with gradual switching between the steps. These observations bring us to the same conclusion on the suitability of this process for neuromorphic device realizations, but for different reasons. From the kinetic point of view, we believe that during PDA the preferred Ag diffusion (i.e., filament formation) pathways are created in the HZO layer, such as GB and TJ, whose diffusion coefficients are typically several orders of magnitude higher than in the bulk of the material \cite{petelin_triple_2001}. Poorer adhesion of unannealed Ag to HZO and formation of diffusion barriers \cite{grisafe_performance_2019} during PDA are also among possible explanations for the features of threshold switching in this case.

\begin{figure}[t!]
    \centering
    \includegraphics[width=\linewidth]{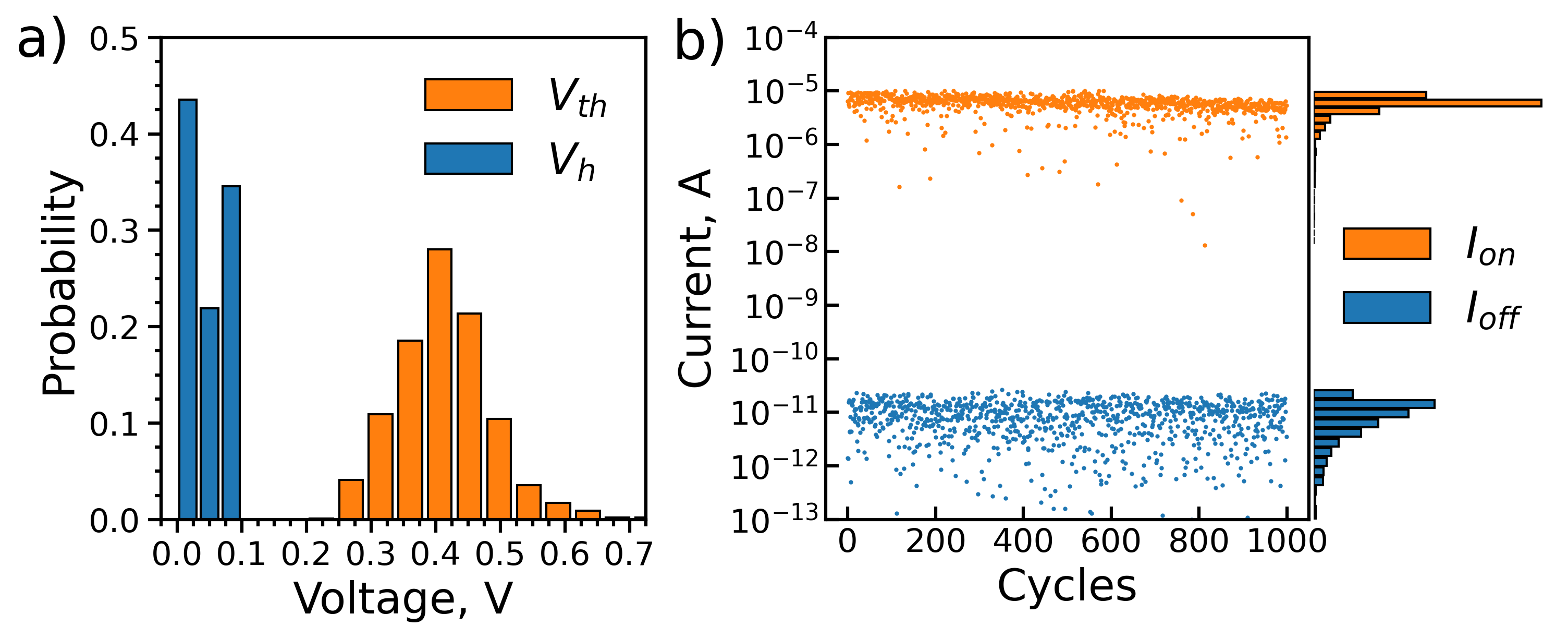}
    \caption{Statistics of (a) switching voltages and (b) currents in ON and OFF states, extracted from 1000 consecutive $I(V)$ cycles, measured on a memristor fabricated in an optimized process (PMA + PDA).}
    \label{fig:statistics}
\end{figure}

{\noindent}The experiments described above show that the devices fabricated with PDA demonstrate a better $I_\text{on}/I_\text{off}$ ratio compared to PMA, but at the same time higher $V_\text{th}$. In order to improve the performance of the threshold switches, two methods were further combined. The first annealing ($550~^{\circ}$C during 30 seconds) is aimed at crystallization of the HZO layer to create preferred pathways for filament formation, similarly to PDA. The second annealing ($350~^{\circ}$C during 60 seconds) is performed after the Ag deposition and is required to facilitate the Ag diffusion through the GB and TJ in HZO without changing the layer's crystallinity.

{\noindent}Indeed, the "PDA + PMA" structures demonstrate the best results in electrical measurements (fig.\ref{fig:IV}). Average $I_\text{on}/I_\text{off}$ ratio is around $10^6$ with SET occurring mostly in a single step. In addition, a low $V_\text{th}$ value of $\approx 0.5~$V was achieved. Thus, the combination of advantages given by two RTA steps allows to obtain the threshold switches that are perfectly suitable for artificial neuron implementation.

{\noindent}In order to count on stable neuromorphic functioning, it is important to verify the reproducibility of the device operation. We have extracted the statistical information on the electrical properties of an optimized "PDA + PMA" device from 1000 consecutive $I(V)$ cycles (\textbf{fig.\ref{fig:statistics}}). It can be seen that while the values of $I_\text{on}$, $I_\text{off}$ and $V_\text{h}$ (hold voltage -- voltage at which switching from ON to OFF occurs) are very stable, while the dispersion of $V_\text{th}$ is relatively larger.

\begin{figure}[t!]
    \centering
    \includegraphics[width=\linewidth]{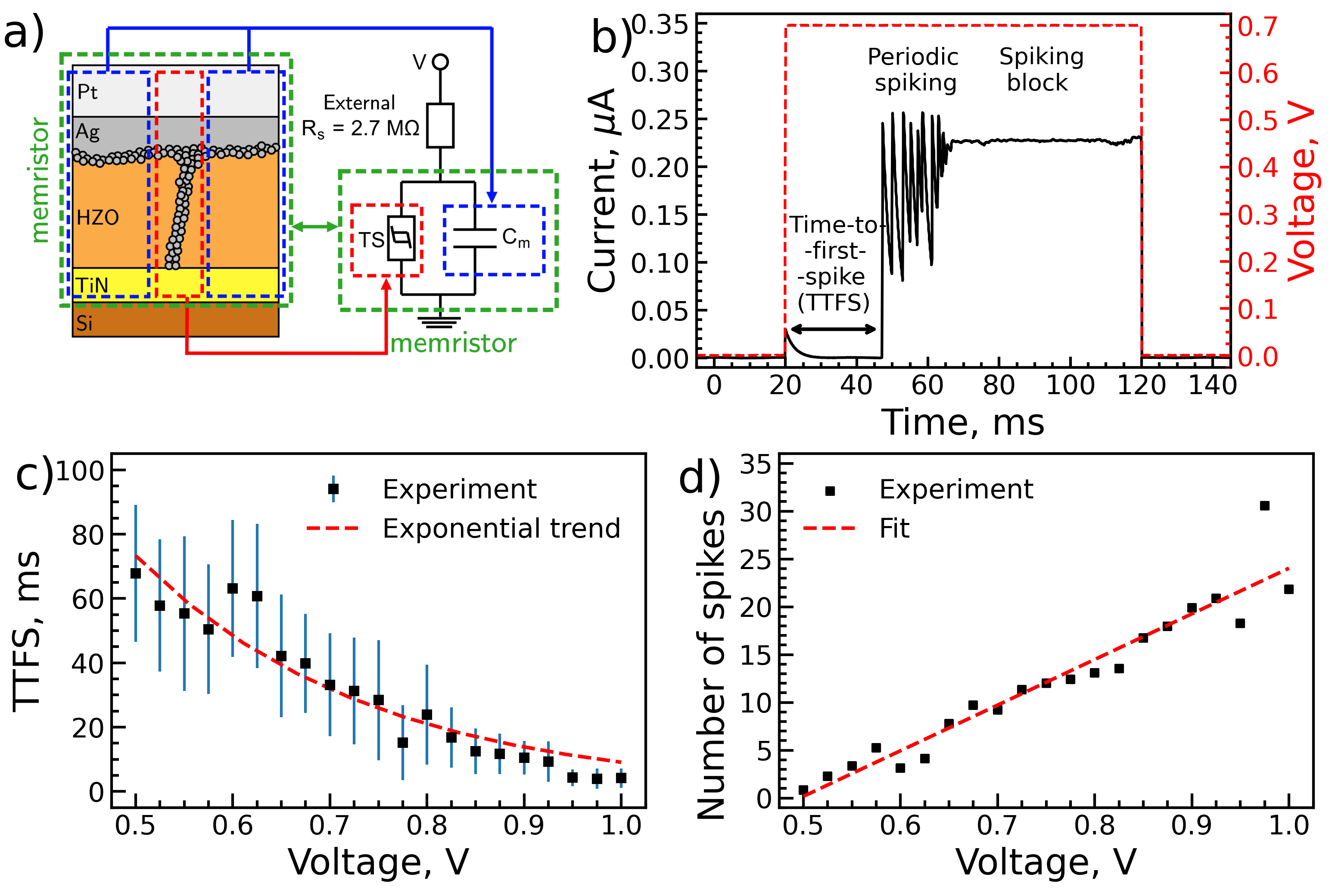}
    \caption{a) Sketch of the filamentary threshold switch and capacitive regions of a Ag/HZO memristor and their correspondence to the elements of a leaky integrate-and-fire neuron circuit. Series resistance $R_{\text{s}}=2.7~$M$\Omega$ was connected externally, while "membrane" capacitance $C_{\text{m}}$ arises naturally from the capacitance of the Ag/HZO/TiN structure. b) Typical current response of a Ag/HZO memristor to a 100$~$ms long rectangular voltage pulse with three characteristic regions: region in the OFF state, which can be used for time-to-first-spike (TTFS) coding; region of relaxation oscillations, corresponding to periodic spiking; region in the ON state, corresponding to spiking block. c) Dependence of TTFS on the voltage value of the input pulse (black squares with blue error bars, calculated statistically from a series of experiments). Red dashed line shows exponential line of trend as a guide for the eye. d) Dependence of the number of spikes in the periodic spiking region on the voltage value of the input pulse (black squares). Red dashed line shows linear fit.}
    \label{fig:neuron}
\end{figure}

Since filamentary switching occupies a small part of the total Ag/HZO/TiN device area ($\sim10^{-4}\div10^{-2}~\upmu$m$^2$ vs. $\sim10^{2}\div10^{4}~\upmu$m$^2$ in our case), our memristor can be considered as a threshold switch and a capacitor connected in parallel (\textbf{fig.\ref{fig:neuron}a}), which already is a realization of a leaky integrate-and-fire (LIF) neuron (a resistor, responsible for charge leakage, is built in the threshold switch in the form of a finite $R_\text{off}$). To stabilize the performance of a memristor by limiting the current flowing through it, we externally connect a resistor $R_{\text{s}}=2.7~$M$\Omega$ in series. Thus, in the following electrical measurements, voltage is applied to one terminal of the resistor $R_{\text{s}}$, while the BE of a memristor is grounded, as indicated on the circuit in fig.\ref{fig:neuron}a.

{\noindent}A unique neuronal functionality is observed in our memristors after the application of constant voltage (rectangular) pulses (\textbf{fig.\ref{fig:neuron}b}). An initial current spike right at the moment of the voltage onset corresponds to the transient process of an RC-circuit with the time constant $\tau \approx 2.5~$ms. Next, current through the device is again close to 0, indicating the retention of the OFF state of the memristor. However, after the period of $\sim1\div100~$ms, we start observing current oscillations, which end with current saturation in the ON state of the memristors. These phenomena can be interpreted in terms of spiking neuron behavior. The initial "idle" period before the current oscillations onset can be used for time-to-first-spike (TTFS) information coding. Moreover, we see clear inverse dependence of TTFS on the input voltage value (\textbf{fig.\ref{fig:neuron}c}), which allows fine output tuning in this neuronal mode. Current oscillations as well can be tuned by the input voltage value to produce different number of spikes at the neuron output with a clear linear dependence (\textbf{fig.\ref{fig:neuron}d}).

\begin{figure}[t!]
    \centering
    \includegraphics[width=\linewidth]{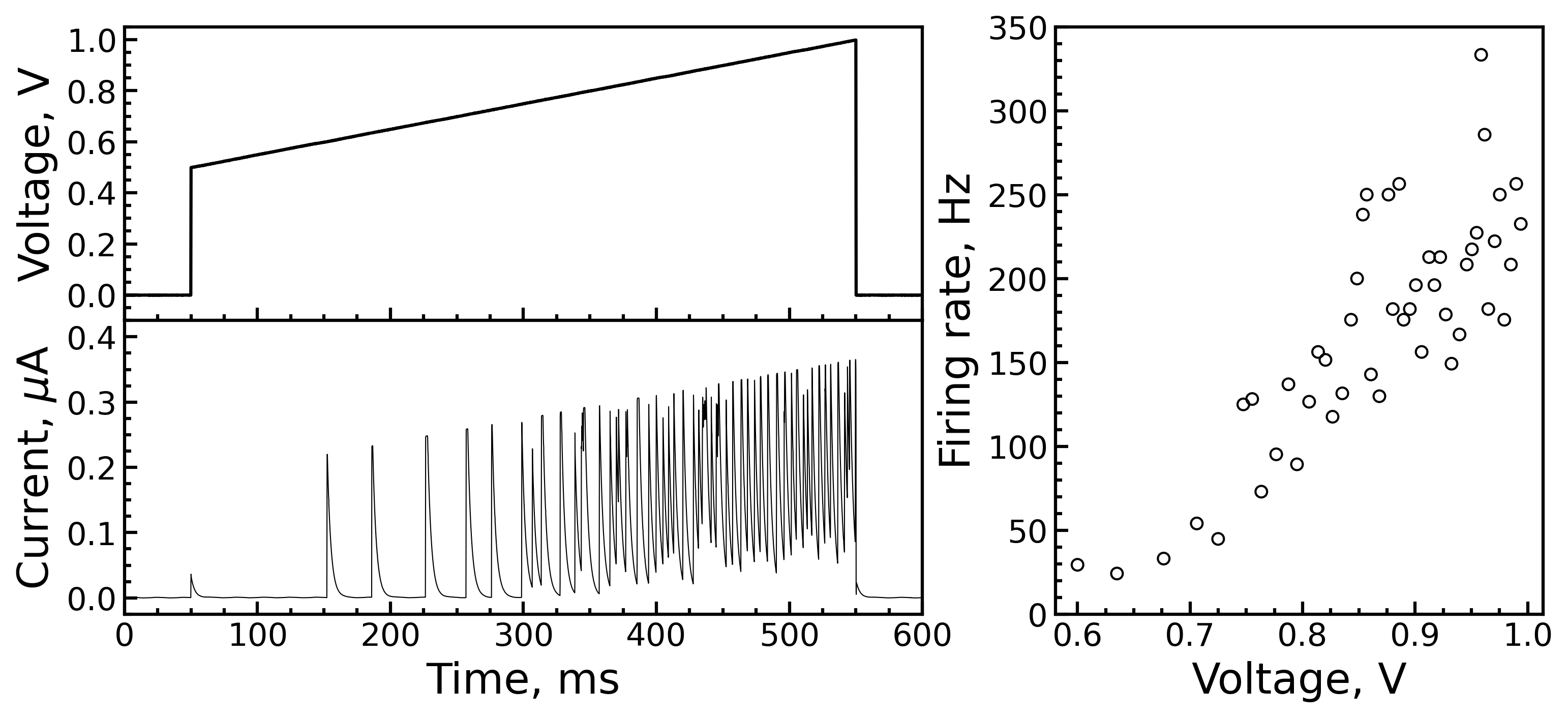}
    \caption{Spiking with variable firing rate in a Ag/HZO memristor via voltage value change.}
    \label{fig:firing_rate}
\end{figure}

{\noindent}In the next experiment, we applied long voltage pulses with a trapezoid profile: voltage value linearly increased with time from 0.5 to 1 V (\textbf{fig.\ref{fig:firing_rate}}). Here, we again see the transient RC spike in the beginning, followed by the idle period. However, the periodic spiking, observed in this mode, occurs with variable time interval between the spikes, decreasing with the voltage increase. The extracted interval values can be recalculated into an instantaneous firing rate, which depends on the input voltage according to an almost linear law. This allows tuning the spiking output in frequency by changing the voltage value.

{\noindent}Thus, our Ag/HZO/TiN memristive spiking neuron supports three different coding modes, controllable by input voltage value: TTFS, number of spikes and firing rate. This multi-functionality can be very beneficial during the design of spiking neural network accelerators based on memristors, since different algorithms sometimes require different procedures to encode the numeric inputs into spike sequences, so the possibility to realize this in a single 3-in-1 device ensures a strong advantage in chip area with respect to three separate dedicated cores.

{\noindent}Average energy consumption by our memristive neuron equals 0.7$~$nJ per spike, which is less than in the best Mott memristor by over an order of magnitude \cite{li_crossmodal_2024} and of the same order as for similar metal ions filamentary memristors \cite{zhou_thermally_2023}.

{\noindent}Finally, we would like to address the choice of a 10-nm-thick HZO layer as a dielectric in our memristive neuron. HZO is known for its ferroelectric properties \cite{chouprik_effect_2023}, and ferroelectrics exhibit variable capacitance with a characteristic voltage dependence, which in several recent works was used to implement purely capacitive memory elements \cite{demasius_energy-efficient_2021, zhang_energy_2023, wu_multi-state_2024, chen_enhanced_2025}. Since we have succeeded in utilizing a built-on capacitance of the Ag/HZO memristor as a membrane capacitance in LIF neuron circuit, it should be possible to realize a reconfigurable neuron with additional rate tunability, provided the ferroelectric degree of freedom is employed. This concept is a subject of our future experiments and, therefore, falls out of scope of this work.

To summarize, in this work we have demonstrated a Ag/HZO memristor-based LIF neuron with multiple spiking functionalities. First, we have optimized the fabrication process of Ag/HZO/TiN memristors by adding two annealing steps (PDA -- after HZO growth, and PMA -- after Ag deposition), which allowed to obtain threshold switches with excellent properties. Next, we utilized an equivalence of our memristor to a LIF neuron circuit and implemented a spiking neuron with three modes of spike coding: TTFS, spike number and firing rate, all controlled by the value of input voltage pulses. The demonstrated threshold memristor fabrication process optimization and multifunctional spiking neuron implementation contribute to the research and development of next-generation energy efficient neuromorphic accelerators.


This work was performed using the equipment of the Center of Shared Research Facilities of Moscow Institute of Physics and Technology and supported by the Russian Science Foundation (project no.24-79-00069). 


\section*{Conflict of Interest}

The authors declare no conflict of interest.

\section*{Data Availability Statement}

The data that support the findings of this study are available from the corresponding author upon reasonable request.

\bibliography{references}


\end{document}